\documentstyle[12pt]{article}
\def\sspace{\baselineskip=.16in}
\def\dspace{\baselineskip=.30in}

\begin{document}
\begin{titlepage}
\begin{center}

{\LARGE \bf Topological Defects and Inflation}
\vskip2truecm
G. Lazarides\\
Physics Division\\
School of Technology\\
University of Thessaloniki, Greece\\
\vskip.5truecm
and\\
\vskip.5truecm
Q. Shafi\footnote{Supported in part by DOE Grant Number DE-FG02-91ER40626.}\\
Bartol Research Institute\\
University of Delaware\\
Newark, DE 19716

\vskip2truecm
\sspace

\begin{abstract} In the context of supersymmetric models, we
analyze the production of topological defects at the end of inflation driven
by a conjugate pair of inflaton fields which are non-singlets under the
continuous symmetry group of the theory.  We find that magnetic monopoles of
mass on the order of $10^{13}\ GeV$ can survive inflation and be present in our
galaxy at an observable level.  We also consider cosmic strings as well as
domain walls.
\end{abstract}

\end{center}
\end{titlepage}

\newpage
\dspace

The new and the chaotic inflationary scenarios [1] typically invoke a very
weakly coupled scalar field known as the inflaton.  The extremely weak
couplings of this field are necessitated by the small value of the observed
temperature fluctuations in the cosmic background radiation. Within the
context of ordinary grand unified theories (GUTS), the inflaton is required to
be a gauge singlet field.  This avoids the `strong' radiative corrections that
the gauge interactions would otherwise produce.  Supersymmetric (SUSY) GUTS
provide us with the possibility of using as inflaton even a gauge non-singlet
weakly coupled field [2].  The idea is to utilize the very same conjugate pair
of standard model (SM) singlet superfields which is also responsible for the
breaking of the gauge symmetry of the theory.  Mutual cancellation of the
D-terms of these conjugate fields can then be easily achieved along some
direction and guarantees the absence of strong radiative corrections which can
spoil the smallness of higgs self-couplings in this D-flat direction.

The intriguing idea that the inflaton is a gauge non-singlet field raises the
interesting possibility that topological defects like magnetic monopoles or
cosmic strings can be produced at the end of inflation.  In ref. 2, we
presented preliminary estimates of topological defect production by a gauge
non-singlet inflaton in the context of a particular class of superstring
theories.  The purpose of this paper is to give, in the context of SUSY GUTS,
a detailed analysis of topological defect production at the end of inflation
driven by a conjugate pair of superfields which are non-singlets under the
continuous (gauge and global) symmetry group of the theory.

We consider a SUSY GUT based on a gauge group G.  The theory possibly
possesses some global symmetries too which may include both continuous and
discrete parts.  Symmetry breaking is obtained through a superpotential which
includes the terms

\begin{equation} W = \kappa S ( -M^2 \ + \ \phi \bar{\phi})_{\cdot}
\end{equation}
Here $\phi, \bar{\phi}$ is a conjugate pair of left-handed SM singlet
superfields which belong to non-trivial representations of the continuous
(gauge and global) symmetry group of the theory and reduce its rank by their
vacuum expectation values (vevs), {\it S\/} is a gauge singlet left-handed
superfield, {\it M\/} is a superheavy mass scale and $\kappa$ a positive
coupling constant.  The superpotential terms in eq. (1) are the only
renormalizable couplings which involve the superfields $S, \phi, \bar{\phi}$
and are consistent with a continuous R-symmetry under which $W \rightarrow
e^{i \theta} W, \ S \rightarrow e^{i \theta} S$ and $\phi \bar{\phi}
\rightarrow \phi \bar{\phi}$.  The non-renormalizable terms $S(\phi
\bar{\phi})^n\ (n \geq 2)$ which are also allowed by this symmetry are assumed
to be small and will be ignored in our discussion.  Moreover, we assume that
the presence of other SM singlets in the theory does not affect the
superpotential in eq. (1) which is responsible for the non-zero vevs of $\phi,
\bar{\phi}$.  The potential obtained from {\it W\/} in eq. (1) is

\begin{equation}
V = \kappa^2 \mid M^2 - \phi \bar{\phi} \mid^2 \ + \ \kappa^2 \mid S \mid^2
(\mid
\phi \mid^2 + \mid \bar{\phi} \mid^2) \ +\ D-terms,
\end{equation}
where the scalar components of the superfields are denoted by the same symbols
as the corresponding superfields.  Vanishing of the D-terms is achieved with
$\mid\bar{\phi}\mid = \mid\phi\mid$ (D-flatness condition).  The
supersymmetric vacuum

\begin{equation}
<S> = 0,\ <\phi> <\bar{\phi}> = M^2,\ \mid <\bar{\phi}> \mid = \mid <\phi> \mid
\end{equation}
lies on the D-flat direction $\bar{\phi}^* = \phi$.  Restricting ourselves to
this particular direction and performing appropriate continuous R- and
non-R-transformations, we can bring the complex $S, \phi, \bar{\phi}$ fields
on the real axis, i.e., $S \equiv {\frac{\sigma}{\sqrt{2}}},\ \bar{\phi} = \phi
\equiv {\frac{1}{2}} \chi$, where $\sigma$ and $\chi$ are normalized real
scalar
fields.  The potential in eq. (2) then takes the form

\begin{equation} V(\chi, \sigma) \ = \ \kappa^2 \left( M^2- \frac{\chi^2}{4}
\right)^2 \ + \ \frac{\kappa^2 \sigma^2 \chi^2}{4} \end{equation} and the
supersymmetric minima become

\begin{equation}
<\chi>=\pm2M \ \ \ , \ \ \ <\sigma>=0_{\cdot}
\end{equation}
It should be noted that inflation and topological defect production with
potentials of the type in eq. (4) were first studied in ref. 3.  However,
these studies were restricted to non-supersymmetric models and used initial
field configurations different than the ones considered here.

The vacuum manifold of the theory, which is obtained from the minimum in eq.
(5) by performing all possible gauge and global transformations, may have
non-trivial homotopical properties in which case the theory predicts the
existence of topological defects.

Following the philosophy of the chaotic inflationary scenario [1], we suppose
that at a cosmic time $t_P \equiv M^{-1}_P$, where $M_P = 1.2 \times
10^{19}$ GeV is the Planck mass, the universe emerges with energy density of
order $M^4_P$ and we concentrate on a particular region of size of order $t_P$
where $\chi$ and $\sigma$ happen to be almost uniform with $\mid\chi \mid \gg
\mid \sigma \mid$.  The potential in eq. (4) is then initially dominated by
the $\chi^4$ term and the initial equations of motion for the $\chi$ and
$\sigma$ fields read

\begin{equation}
\ddot{\chi} \ + \ 3H \dot{\chi} \ + \ \frac{\kappa^2 \chi^3}{4} \  \simeq \ 0
\end{equation}
and
\begin{equation} \ddot{\sigma} \ + \ 3H\dot{\sigma} \ + \  \frac{\kappa^2
\chi^2 \sigma}{2} \ \simeq \ 0, \end{equation}
where
\begin{equation}
H \ = \ \left( \frac{8 \pi}{3} \right)^{1/2} M^{-1}_P \rho^{1/2} \ = \ \left(
\frac{8
\pi}{3} \right)^{1/2} M^{-1}_P \left( \frac{1}{2} \dot{\chi}^2 \ + \
\frac{1}{2} \,
\dot{\sigma}^2 \ + \ V(\chi, \sigma) \right)^{1/2}
\end{equation}
is the Hubble parameter, the dots denote derivatives with respect to cosmic
time and $\rho$ is the energy density.  It is easily seen that for $\chi \gg
M_P/(3 \pi)^{1/2}$, eqs. (6) and (7) reduce to

\begin{equation}
3H \dot{\chi} \ + \ \frac{\kappa^2 \chi^3}{4} \ \simeq \ 0
\end{equation}
and

\begin{equation}
3H \dot{\sigma} \ + \ \frac{\kappa^2 \chi^2 \sigma}{2} \ \simeq \ 0
\end{equation}
respectively, while the kinetic terms of the $\chi$ and $\sigma$ fields in the
RHS of eq. (8) can be neglected.  During this inflationary period, eqs. (9)
and (10) give

\begin{equation}
\chi \ = \ \chi_o \ exp \left(- \frac{\kappa M_P \Delta t}{(24 \pi)^{1/2}}
\right)
\end{equation}
and

\begin{equation} \sigma \ = \ \sigma_o \ exp \left(- \frac{2\kappa M_P \Delta
t}{(24 \pi)^{1/2}} \right), \end{equation} where $\chi_o$ and $\sigma_o$ are
the initial values of the fields at time $t_P (\chi_o \gg \sigma_o)$ and
$\Delta t \equiv t-t_P$.  Eqs. (11) and (12) imply that $\chi^2/\sigma \ = \
\chi^2_o/\sigma_o$ and the ratio $\sigma / \chi$ decreases exponentially with
cosmic time.  From eq. (9), the number of e-foldings, $N(\chi)$, from when the
field has value $\chi$ till inflation ends turns out to be

\begin{equation}
N(\chi) \ = \ \pi \ \frac{\chi^2}{M_P^2} \ .
\end{equation}

The contribution of the scalar metric perturbation to the microwave background
quadrupole anisotropy (scalar Sachs-Wolfe effect) is given by [4]

\begin{equation}  \left( \frac{\Delta
T}{T} \right)_S \ \simeq \ \left.\left( \frac{32 \pi}{45} \right)^{1/2} \
\frac{V^{3/2}}{M^3_P \left( \partial V / \partial \chi \right)} \ \right|_{k
\sim H} \ = \left.\left( \frac{32\pi}{45} \right)^{1/2} \ \frac{\kappa
\chi^3}{16
M^3_P} \right|_{k \sim H}, \end{equation}
where the right-hand side is evaluated at the value of the $\chi$ field where
the length scale $k^{-1}$, which corresponds to the present horizon size,
crossed outside the de Sitter horizon during inflation.  Substituting $\chi$
from eq. (13) in eq. (14), we obtain

\begin{equation}
\left( \frac{\Delta T}{T} \right)_S \ \simeq \ \left( \frac{32}{45}
\right)^{1/2} \
\frac{\kappa}{16 \pi} \ N^{3/2}_H,
\end{equation}
where $N_H$ is the number of e-foldings of the present horizon size during
inflation.  The gravitational wave contribution $\left( \frac{\Delta T}{T}
\right)_T$ to the quadrupole anisotropy is [5]

\begin{equation}
\left( \frac{\Delta T}{T} \right)_T \ \simeq \ 0.78\frac{V^{1/2}}{M^2_P} \ = \
0.78 \frac{\kappa}{4 \pi} N_{H \cdot}
\end{equation}
Taking $N_H = 55$ and $(\Delta T/T) \simeq 7 \times 10^{-6}$ from COBE, we
then obtain $\kappa \simeq 0.92 \times 10^{-6}$ and $r \equiv \left(
\frac{\Delta
T}{T}\right)^2_T / \left( \frac{\Delta T}{T} \right)^2_S \ \simeq \ 0.25$.
The spectral index

\begin{equation}
n \ = \ 1 - \frac{3}{N_H} \ \simeq \ 0.945
\end{equation}
turns out to be very close to the Harrison-Zeldovich value $(n=1)$ and lies
in the central range of values preferred by observations.

At the end of inflation at cosmic time $t_f \ \sim \ H^{-1}_f \ \sim \ 3(6
\pi)^{1/2} (\kappa M_P)^{-1}$, the $\chi$ field starts performing damped
oscillations over its maximum at $\chi=0$ with frequency of order $\kappa
\chi_m$, where $\chi_m$ is the amplitude of these oscillations.  As is well
known [6], an oscillating field with $\chi^4$ potential behaves like
radiation, i.e., $\rho + p \ = \ (4/3) \rho$, where $p$ is the pressure
averaged over one oscillation time of this field.  From eq. (8) and  the
continuity equation $\dot{\rho} = -3H(\rho+p)$, we obtain $H
\simeq 1/2t$ and $\chi_m \ \simeq \ (3/2 \pi)^{1/4}\ (M_P/\kappa t)^{1/2}$.
After inflation is over, eq. (7) averaged over one oscillation of $\chi$ takes
the form

\begin{equation}
\ddot{\sigma} \ + \ \frac{3}{2t} \dot{\sigma} \ + \ \left( \frac{3}{32 \pi}
\right)^{1/2} \ \frac{\kappa M_P}{t} \sigma \ \simeq \ 0_{\cdot}
\end{equation}
For $t \gg t_f$, the `frequency' of the $\sigma$ field is of order

\begin{equation} \left( \frac{3}{32 \pi}\right)^{1/4} \ \left( \frac{\kappa
M_P}{t} \right)^{1/2} \end{equation} and is much greater than $H \simeq
1/2t$.  Thus, $\sigma$ also starts performing damped oscillations about $\sigma
= 0$
but with an initial amplitude much smaller than the amplitude of $\chi$.  The
presence of the $\sigma$ field is not essential for our
subsequent arguments and, for simplicity, we will put it equal to zero
in the rest of the discussion.

Topological defects associated with the symmetry breaking caused by the
non-zero vev of the $\chi$ field assumed to be a gauge non-singlet can, in
principle, be produced at cosmic time $t_d \sim H^{-1}_d \ = \ (3/8
\pi)^{1/2}\ (M_P/\kappa M^2)$, where the energy density $\rho$ of the
oscillating $\chi$ field reduces to the value $\kappa^2 M^4$.  At this point
the $\chi$ field ceases to oscillate over the potential barrier $V(\chi=0,\
\sigma=0) \ =\ \kappa^2 M^4$ at $\chi=0$ and gets trapped in one or the other
of the potential wells associated with the two minima in eq. (5).  As we
explained earlier, the oscillations of the $\chi$ field, for cosmic times
between $t_f$ and $t_d$, are approximately governed by a $\chi^4$ potential
and the system behaves almost like radiation.  This means that any density
fluctuations at scales smaller than the horizon at $t_d$ are erased.  So, at
cosmic time $t_d$, the $\chi$ field  is expected to be uniform on scales
smaller than about $t_d$ and therefore fall into the same potential well
everywhere within a particle horizon.  Thus, the smallest possible distance
between neighbouring defects produced at $t_d$ is of order $t_d$. This maximal
number of topological defects per horizon at $t_d$ is achieved [7] only if the
inflationary density fluctuation on scale $t_d$, $(\delta \rho / \rho)_d$,
exceeds the fraction of energy density lost, $(\delta \rho_{1/2}/ \rho)$,
within half a cycle of oscillation
of the $\chi$ field at $t_d$.  This energy loss
can be calculated by using the equation of continuity $\dot{\rho} = -3H
\dot{\chi}^2$ and eqs. (8) and (4),

\begin{equation}
\begin{array}{lcl}
\frac{\delta \rho_{1/2}}{\rho} \ & = & \ \frac{6 H}{\kappa^2 M^4} \ \int^{2
\sqrt{2} M}_o \ \dot{\chi}d\chi \ = \ \frac{6H}{\kappa^2 M^4} \ \int^{2
\sqrt{2} M}_o \ \sqrt{2(\rho-V)} \ d\chi \\
\\
& = & \ 16\left( \frac{8 \pi}{3} \right)^{1/2} \ \frac{M}{M_P} \ .
\end{array}
\end{equation}
Moreover, the inflationary density fluctuation is

\begin{equation}
\left( \frac{\delta\rho}{\rho} \right)_d \ \simeq \ \left(
\frac{\delta\rho}{\rho} \right)_H \ \left( \frac{N_d}{N_H} \right)^{3/2},
\end{equation}
where [1]

\begin{equation}
\left( \frac{\delta\rho}{\rho} \right)_H \ \simeq \ 9.3 \left( \frac{\Delta
T}{T} \right)_S \ \simeq \ 5.8 \times10^{-5}
\end{equation}
is the density perturbation on the present horizon scale and $N_d$ is
the number of e-foldings which the particle horizon at $t_d$ suffered during
inflation.  To estimate $N_d$, recall that between $t_f$ and $t_d$ our system
behaves like radiation and the scale factor of the universe increases by
a factor $(V(\chi = \chi_f)/\kappa^2M^4)^{1/4} \ = \ \chi_f/2M \ = \
(12\pi)^{-1/2}(M_P/M)$.  The horizon size at $t_d$ at the end of inflation
is $(3/\sqrt{2})(\kappa M)^{-1}$. Comparing this with $H^{-1}_f$ we find
$exp(N_d) \ \simeq \ (12\pi)^{-1/2}(M_P/M)$.  Eqs. (20), (21), (22) and the
condition for `maximal' production of topological defects at $t_d$ then imply
that $M \leq 1.9 \times 10^{12}$ GeV or $<\chi> \ \leq \ 3.8 \times 10^{12}$
GeV.

Let us first consider the case of magnetic monopoles.  Their initial number
density, if they are `maximally' produced at $t_d$, is expected to be $n_M
\sim t^{-3}_d$ and so the relative initial energy density in monopoles is

\begin{equation} d_M \ = \ \frac{n_M m_M}{\kappa^2 M^4} \ \sim \ \left(
\frac{8 \pi}{3} \right)^{3/2} \ \frac{\kappa M^2 m_M}{M^3_P}, \end{equation}
where $m_M$ is the monopole mass.  At cosmic times greater than $t_d$, the
$\chi$ field performs damped coherent oscillations about the minima in eq.
(5).  The potential is now approximately quadratic, the system behaves like
matter and so $d_M$ remains constant.  The oscillating inflaton will
eventually decay into lighter particles and `reheat' the universe to
temperature $T_r$. The process of `reheating' is strongly dependent on the
particle physics model one adopts, but bounds can be obtained in a relatively
model independent way.  Assume that the inflaton field, which has mass
$m_{\chi} = \sqrt{2} \kappa M$, decays to a pair of particles (say fermions)
with mass $m \leq m_{\chi}/2$.  The relevant effective coupling constant $f
\leq m/<\chi>$.  The decay rate $\Gamma \ \sim \ f^2 m_{\chi} \ \leq \ (1/4
\sqrt{2})\kappa^3 M$ and the `reheat' temperature $T_r \ \sim \ (\Gamma
M_P)^{1/2} \ \leq \ 2^{-5/4} \kappa^{3/2}(MM_P)^{1/2}$.  At `reheat', $d_M \
\simeq \ n_M m_M/sT_r$, where $s$ is the entropy density and, from eq. (23),

\begin{equation}
\frac{n_M}{s} \ \simeq \ \left( \frac{8 \pi}{3} \right)^{3/2} \ \frac{\kappa
M^2 T_r}{M^3_P} \ \leq \ 2^{-5/4} \left( \frac{8 \pi}{3} \right)^{3/2} \left(
\frac{\kappa M}{M_P} \right)^{5/2} \ \leq \ 10^{-31}\ .
\end{equation}

Thus, a flux of intermediate mass ($\sim 10^{13}$ GeV) magnetic monopoles close
to the Parker bound may exist in our galaxy.

Turning now to cosmic strings, the string network, which enters the particle
horizon at some cosmic time subsequently approaches the well-known scaling
solution [8].  The requirement that a string network exists in the present
universe  means that the scale of this network at production time  is greater
than or equal to $t_d$, and must not exceed the scale of the present universe
at $t_d$.  This is achieved if $(\delta \rho/\rho)_H \ \stackrel{_>}{_\sim} \
\delta \rho_{1/2}/\rho$, which gives $M \ \leq 1.5 \ \times \ 10^{13}$ GeV or
$<\chi> \ \leq \ 3 \times 10^{13}$ GeV.  These strings presumably are too
`light' to play any role in galaxy formation.

Domain walls are cosmologically catastrophic and must be avoided.  Their case
is very similar to the cosmic string case we have just discussed and one
finds that the problem is avoided if $<\chi> \ \geq \ 3
\times 10^{13}$ GeV.  So, if the non-zero vev of the inflaton $\chi$ breaks
some discrete symmetries, we should make sure that this vev is of `superheavy'
($\gg 10^{12}$ GeV) scale.

Our analysis can be extended to the inflationary scenario of ref. 2 where the
inflaton $\phi$, being a gauge non-singlet, is responsible for the `radiative'
breaking of the $SU(3)^3$ gauge symmetry of the theory.  The relevant
potential consists of a negative $\rm{mass}^2$ term characterized by the
supersymmetry breaking scale $M_s \ \sim \ \rm{TeV}$, and a non-renormalizable
term $(\phi^6/M^2_P)$ with a small dimensionless coupling fixed by $\delta
\rho / \rho$.  The  the vev of $\phi$ is determined to be on the order of the
GUT scale so that topological defects do not arise.

In summary, in the framework of supersymmetric models, we estimated the
density of topological defects that can be produced at the end of an
inflationary epoch driven by a conjugate pair of gauge non singlet fields.  We
find that magnetic monopoles close to the Parker bound and/or cosmic
strings may be produced if the inflaton vev takes intermediate scale values.

\end{document}